\documentclass[]{interact}

\usepackage{epstopdf}
\usepackage[caption=false]{subfig}
\usepackage[doublespacing]{setspace}
\usepackage[natbibapa,nodoi]{apacite}
\usepackage{graphicx}
\usepackage{color}
\usepackage{amsmath}
\usepackage{amssymb}
\usepackage{amsfonts}
\usepackage[normalem]{ulem}
\usepackage{multirow}
\usepackage{bigstrut}
\usepackage{booktabs}
\usepackage{tabularx}
\usepackage{bm}

\def\eqx"#1"{{\label{#1}}}
\def\eqn"#1"{{\ref{#1}}}

\def\var{\mbox{var}}

\def\pmatrix#1{\left(\begin{array}{#1}}
\def\endpmatrix{\end{array}\right)}

\newcommand{\bc}{\color{black}}

\newcommand{\ec}{\color{black}}

\newcommand{\bfx}{ {\bf x} }

\newcommand{\bfY}{ {\bf Y} }
\newcommand{\bfe}{ {\bf e} }

\newcommand{\bfSigma}{\mbox{\boldmath $\Sigma$}}
\newcommand{\bfbeta}{\mbox{\boldmath $\beta$}}

\newcommand{\hbfbeta}{\widehat {\mbox{\boldmath $\beta$}}}
\newcommand{\hbfSigma}{\widehat {\mbox{\boldmath $\Sigma$}}}

\newcommand{\hbfe}{\widehat{\bf e}}
\newcommand{\hbfr}{\widehat{\bf r}}

\theoremstyle{plain}

\theoremstyle{definition}

\theoremstyle{remark}

\begin{document}

\articletype{}

\title{Permutation tests for assessing potential non-linear associations between treatment use and multivariate clinical outcomes}

\author{
\name{Boyu Ren\textsuperscript{a}, Stuart R. Lipsitz\textsuperscript{b}\thanks{CONTACT Stuart R. Lipsitz. Email:  slipsitz@bwh.harvard.edu}, Garrett M. Fitzmaurice\textsuperscript{a} and Roger D. Weiss\textsuperscript{a}}
\affil{\textsuperscript{a}McLean Hospital, Belmont, MA, U.S.A.; \textsuperscript{b} Brigham and Women’s Hospital and Ariadne Labs, Boston, MA, U.S.A.}
}

\maketitle

\begin{abstract}

In many psychometric applications, the relationship between the mean of an outcome and a quantitative covariate is too complex to be described by simple parametric functions; instead, flexible nonlinear relationships can be incorporated using penalized splines. Penalized splines can be conveniently represented as a linear mixed effects model (LMM), where the coefficients of the spline basis functions are random effects. The LMM representation of penalized splines makes the extension to multivariate outcomes relatively straightforward. In the LMM, no effect of the quantitative covariate on the outcome corresponds to the null hypothesis that a fixed effect and a variance component are both zero. Under the null, the usual asymptotic chi-square distribution of the likelihood ratio test for the variance component does not hold. Therefore, we propose three permutation tests for the likelihood ratio test statistic: one based on permuting the quantitative covariate, the other two based on permuting residuals. We compare via simulation the Type I error rate and power of the three permutation tests obtained from joint models for multiple outcomes, as well as a commonly used parametric test. The tests are illustrated using data from a stimulant use disorder psychosocial clinical trial.
\end{abstract}


\section{Introduction}

In many psychosocial clinical studies, subjects' health and well-being are
assessed in terms of multiple outcomes. A common analytic goal is to assess the
functional relationship between each of these outcomes and a quantitative
covariate. For example, our motivating application is from a longitudinal
clinical trial, the National Institute on Drug Abuse Collaborative Cocaine
Treatment Study (CCTS) \citep{crits1999psychosocial}, where there is scientific
interest in 6 month post-treatment changes in 5 psychosocial problem domains.
Thus, the outcomes are the simple change scores in legal, employment, family,
psychological and medical problems as determined by the difference in baseline
and 6 month post-treatment follow-up assessments using the 5 domain scores from
the Addiction Severity Index (ASI).  The main covariate of interest is a
quantitative measure of within-treatment frequency of use of cocaine; this  was
defined as the proportion of positive monthly urine toxicology screens during
the 6 month duration of treatment. To assess the relationship between each of
these 5 outcomes and the quantitative measure of within-treatment use of
cocaine, we can fit linear regression models for each of the five change scores
as a function of this quantitative covariate. \bc However, since these outcomes are likely to be correlated, a joint model for them has the potential to yield more powerful tests of the treatment effects compared to the separate linear models (for example, see \citealp{yoon2011alternative}). \ec To this end, one can use a
linear mixed effects model to account for the correlation among the multiple outcomes on the same study participant. 

Because the relationships
between the quantitative covariate (within-treatment use) and changes in the 5
problem domains may not be linear, we propose fitting flexible 'splines' or
piecewise linear relationships that allow the data to determine the form of the
relationships. In particular, we use penalized splines that can be represented 
within linear mixed effects models by regarding the coefficients of the basis
functions for the splines as random effects from a normal distribution with a
single variance component. The linear mixed effects representation of penalized
splines is straightforward to incorporate in a linear mixed effects model for
multivariate outcomes. 

{ Typically, the linear mixed effects representation of penalized piecewise linear splines for a univariate regression
has a fixed effect for the linear effect of the quantitative covariate and
random effects for the coefficients of the basis functions for the spline, where
the random effects are assumed to come from a single normal distribution with
common (unknown) variance component.  In this paper,  we consider the case of multivariate outcomes, where we have (say) $K$ outcomes
per subject, and our linear mixed effects model allows different penalized
splines for the relationship between the quantitative covariate and each of
the $K$ outcomes,  leading to a linear mixed model with $K$ variance components
for the penalized splines, as well as $K$ fixed effects for the linear terms of
the quantitative covariate.  In particular,  our interest is in the joint test
that all of the $2K$ parameters (the fixed effects and penalized spline variance
components) for the associations between the quantitative covariate  and $K$
outcomes are 0. That is, a joint test of zero-effect of the covariate on the $K$ outcomes. \bc In general, for testing that a variance component equals 0, it is well-known that the usual asymptotic chi-square distribution of the
likelihood ratio (LR) test under the null does not hold \citep{miller1977asymptotic,lin1997variance,verbeke2003use}. This provides the motivation for considering permutation tests as a practical alternative.}
\ec


{ In this paper,  we propose three permutation tests for the joint null
hypothesis that the $K$ random effects spline variance component, as well as the
$K$ fixed linear effects for the quantitative covariate, are all equal to zero.
Previous work in permutation tests for variance components in linear
mixed effects models has been done by \cite{pesarin2010permutation}, \cite{samuh2012use}, \cite{lee2012permutation}, \cite{drikvandi2013testing}, and \cite{du2020testing}. Particularly relevant to our approach for
penalized splines with a univariate outcome ($K=1$),  \cite{lee2012permutation} used
the linear mixed effects representation of penalized splines to propose a
permutation test formed by permuting residuals.  \cite{lee2012permutation} had first
developed their permutation test for random effects variance components in linear
mixed models (not particularly for penalized splines),  and then showed that
since penalized splines can be represented as a linear mixed model, that their
permutation test for variance components could be applied to penalized splines.
The \cite{lee2012permutation} permutation test for a penalized splines was proposed
for testing  a variance component for a penalized spline with a univariate outcome ($K=1$) equals 0, but not for a
joint test of fixed effects and variance components for multivariate outcomes ($K>1$). }

{ Our three permutation tests for  the joint null hypothesis that the $K$
spline variance component and the $K$ fixed linear effects are equal to zero are: 1)
permutation of the quantitative covariate in a linear mixed model for multiple
outcomes; 2) permutation of the residual vector (under the null) of the $K$ outcomes for a
subject, with residual vectors permuted across subjects;  3) permutation of a Cholesky
transformation of the
residual vector (under the null) of $K$ outcomes for a subject, creating univariate Cholesky
transformed residuals that can be permuted both within and across subjects. 
The third permutation test can be considered an extension of \cite{lee2012permutation} to a joint test of the $K$ fixed effects and $K$ spline variance
components.  As an alternative to the permutation tests,  we also consider the
parametric approach by \cite{wood2013p}  that is commonly
used in practice for testing the significance of penalized spline terms.  \bc Of note, \ec Wood's approach (hereafter referred to as the parametric generalized additive model test, or parametric GAM test) uses a Wald statistic derived from the estimated smoothing components{\bc, which follows a mixture of chi-square distributions.\ec} This approach does not use the random effects representation of the penalized
spline, but optimizes a generalized cross-validation statistic to find a penalty
parameter.  We use simulation to then compare the three permutation approaches,
along with the \cite{wood2013p} approach, under \bc three different numbers ($N$) of subjects
($N=50,100,200$), \ec two different numbers of outcomes ($K=5,10$), and two different
joint distributions (normal,  scaled log-normal) for the multivariate ($K$)
outcomes on
a subject; we also vary the magnitude of the correlation among the $K$ outcomes as well as the relation between the outcome and the covariate of interest. }

\bc
In Section 2, we describe the linear mixed effects model representation of penalized splines for
multivariate outcomes. In Section 3, we describe the three proposed permutation tests of zero-effect of the quantitative covariate on the multiple outcomes. \ec  In Section 4, we present the results of the 
simulation study to assess the Type I error and power of the joint testing
procedures. In univariate regression settings, permutation tests have been found to be robust to non-normal outcomes \citep{WINKLER2014381},  and we explore this in the simulations in Section 4 for multivariate outcomes. In Section 5, the proposed methods are illustrated in
analyses of the data on the 5 problem domains from the Collaborative Cocaine
Treatment Study. 
 
\section{Linear Mixed Model Representation for Penalized Splines}

Let $Y_{ik}$ denote the $k^{th}$ outcome ($k=1,...,K)$  for subject $i$
$(i=1,...,N)$  and let $\bfY_i=(Y_{i1},...,Y_{iK})'$ be the vector of the $K$
outcomes for subject $i$.  Let $\bfx_i$ denote a covariate vector for subject
$i$ and let $s_i$ denote the quantitative covariate of primary interest, i.e., $\bfx_i$ are the covariates besides the primary covariate of interest, $s_i$.  We
assume that $\bfY_i$ given $(\bfx_i,s_i)$ is multivariate normal with marginal
model for each outcome given by
\begin{equation}
Y_{ik} = \beta_{0k} + \bfbeta_{1k}'\bfx_i + g_k(s_i) + e_{ik}
\eqx"LMM1"
\end{equation}
where $(\beta_{0k},\bfbeta_{1k}')$ are unknown regression parameters for outcome
$k,$ $g_k(s_i)$ is an unknown function of interest for the relationship between
the quantitative covariate $s_i$ and outcome $k,$  and $e_{ik} \sim
N(0,\sigma_k^2).$  Because the multiple outcomes need not have the same marginal
normal distribution, { we recommend for most applications that the covariance matrix for the vector of outcomes for a subject is assumed to be unstructured,}
$Var(\bfY_i|\bfx_i,s_i) = \bfSigma$ with $Var(Y_{ik}|\bfx_i,s_i) = Var(e_{ik}) =
\sigma_k^2$ and $Cov(Y_{ij},Y_{ik}|\bfx_i,s_i) = \sigma_{jk},$ \bc for $j \ne k$.\ec

Suppose, for ease of exposition, we model $g_k(s_i)$ as a piecewise-linear
spline with $C$ knots (located at $\kappa_1,
\kappa_2,...,\kappa_C);$ however, we note that the proposed method can also be
generalized to any set of basis functions for the spline (e.g., cubic spline or
B-spline basis; the former uses a cubic polynomial in the interval between successive knots, the latter is an alternative  parameterization that has higher numerical stability).  
Then, we can rewrite (\eqn"LMM1") as
\begin{equation}
Y_{ik} = \beta_{0k} + \bfbeta_{1k}'\bfx_i + \gamma_k s_i + \sum_{c=1}^C a_{ck}
(s_i-\kappa_c)_{+}
+ e_{ik}
\eqx"LMM2"
\end{equation}
where $\gamma_k$ is the unknown fixed effect coefficient of the linear term for
$s_i$ and  the truncated line function $(s_i-\kappa_c)_{+}= (s_i-\kappa_c)$ if
$(s_i-\kappa_c)>0$ and is 0 otherwise, and $a_{ck}$ is the unknown coefficient
for the $c^{th}$ truncated line function for outcome $k.$  Inclusion of the
$(s_i-\kappa_c)_{+}$ terms allows for a piecewise linear relationship with
potentially different slopes between the knot locations, $\kappa_1,
\kappa_2,...,\kappa_C.$  Note, however, that a model with too many knots can
yield a fitted curve that is not very smooth. 
An alternative approach to obtain
a smooth spline curve is to use a penalized spline where a large number of knots
are retained but their influence is constrained by shrinking many of the
$a_{ck}$'s toward zero.  Penalized spline regression is performed by requiring
that the sum of squares
\begin{equation}
 \sum_{c=1}^C a_{ck}^2
\eqx"SSC"
\end{equation}
is less than some chosen positive value for each outcome $k$, \bc referred to as the penalty term. \ec Using a fixed effect model for the $a_{ck}$'s, the penalty term
can be chosen by generalized cross-validation (GCV; \citealp{craven_smoothing_1978}); alternatively,
it can be estimated \bc from the data at hand \ec
using a linear mixed effects model representation of penalized splines.  That
is, the connection with linear mixed effects models also provides an automatic
choice of the amount of smoothing via the estimation of the penalty term as the
ratio of variances in the mixed effects model.  { Intuitively,  the linear
mixed effects model representation of penalized splines is equivalent to putting
a ridge penalty on (\eqn"SSC"), and thus shrinks the $a_{ck}$'s toward zero. }
In particular, penalized splines can be implemented by regarding the
coefficients $a_{1k},...,a_{Ck}$ \bc for the truncated line functions \ec in (\eqn"LMM2") as random effects in
the linear mixed effects model, with independent normal distribution $a_{ck}
\sim N(0,\sigma_{ak}^2)$.  
\bc
That is,  the random effects  
$a_{1k},...,a_{Ck}$ in the penalized spline model for the $k^{th}$ outcome variable  $Y_{ik}$ are assumed to be an iid sample from a normal distribution, $a_{ck}
\sim N(0,\sigma_{ak}^2)$; the magnitude of $\sigma_{ak}^2$ (relative to the error variance, $\sigma_{k}^2$) determines the amount of smoothing for the $k^{th}$ outcome variable. To allow each outcome to have its own smoothing parameter, there is a separate $\sigma_{ak}^2$ ($k=1,...,K$) for each of the $K$ outcome variables.
\ec
{ For the connection between the linear mixed model
representation of penalized splines and the ridge penalty, see for example, \cite{wang1998mixed,wang1998smoothing},  \cite{ruppert2003semiparametric} and \cite{fitzmaurice2012applied}, Chap. 19. }  

\bc
When $a_{1k},...,a_{Ck}$ are treated as random effects, the penalized spline model given by (\eqn"LMM2") is a linear mixed effects model since it models the mean of $Y_{ik}$ in terms of a combination of fixed effects, 
$(\beta_{0k},\bfbeta_{1k}',\gamma_{k})$, and random effects, $a_{1k},...,a_{Ck}$. Although it satisfies the technical definition of a linear mixed effects model, we note that the model given by (\eqn"LMM2") differs from many conventional and widely-used linear mixed effects models in the following two ways. First, the random effects are indexed by $k$, the index for the different outcome variables, and not by $i$, the index for different individuals. As a result, both the fixed effects and the random effects in (\eqn"LMM2") are shared by all individuals. Second, unlike many conventional linear mixed effects models, the random effects in (\eqn"LMM2") are emphatically not considered to be a random sample of levels drawn from some larger $``$population$”$. In particular, it does not make sense to imagine taking more draws from the random effects distribution. That is, we do not think of 
$ \beta_{0k} + \bfbeta_{1k}'\bfx_i + \gamma_k s_i + \sum_{c=1}^C a_{ck}
(s_i-\kappa_c)_{+}$ arising as a draw from a random mechanism; instead, it is thought of as being fixed and unknown. 
The random effects are simply included in the model as a device for smoothing or constraining the magnitudes of the coefficients for the basis functions. The penalty term that determines the amount of smoothing is given by the ratio of the error variance, $\sigma^{2}_{k}$, to the variance of the random effects, $\sigma_{ak}^2$.  When $\sigma_{ak}^2 = \infty$, there is no penalty and the coefficients for the basis functions are unrestricted and can be expected to overfit the data. When $\sigma_{ak}^2$ is finite, there is some amount of smoothing resulting from smaller estimates of 
$a_{ck}$ and corresponding decreases in the influence of the basis functions, $(s_i-\kappa_c)_{+}$. 

\ec 

An appealing feature of this penalized spline mixed effects model is that any
non-linearity in the effect of $s_i$ on $Y_{ik}$ can be determined by testing
the null hypothesis that the variance $(\sigma_{ak}^2)$ of the
$a_{1k},...,a_{Ck}$ is zero; moreover, when $\sigma_{ak}^2$ is not zero, the
nature of the relationship can be determined by obtaining the  ``best
linear unbiased predictor" (BLUP) of the random effects, $a_{1k},...,a_{Ck}$ \bc \citep{henderson1975best}, \ec
i.e., estimates or predictions of $a_{1k},...,a_{Ck}$ from the data at hand (and
model-based estimates of the fixed effects and variance components).  This
linear mixed effects model assumes that we have only a single realization of
$a_{1k},...,a_{Ck}$, and these $C$ random coefficients are shared by all
individuals.  For outcome $k,$ the test of no association between $s_i$ and
$Y_{ik}$, i.e., $\mathbb E(Y_{i,k}|s_i,\bm x_i)$ does not depend on $s_i$, translates into a joint test of $H_0:\gamma_k =\sigma_{ak}^2=0,$  {
for $k=1,...,K.$   In particular,  our interest is in the joint test that all of
the $2K$ parameters (the fixed effects and penalized spline variance components)
for the association between $s_i$ and $Y_{ik}$ are 0. }

The (restricted) maximum likelihood estimate of the parameters
$(\beta_{0k},\bfbeta_{1k}',\gamma_k,\sigma_{ak}^2)$ $(k=1,...,K)$ and $\bfSigma$
is  obtained by maximizing the marginal multivariate normal likelihood; this can be implemented in any linear mixed effects model
software program, e.g., lmer function in R \citep{rman} or PROC MIXED in SAS \citep{SAS-STAT}.  Further,
the BLUP predictions of $a_{1k},...,a_{Ck}$ can also be obtained in such a
program.  Next, consider the test that $\mathbb E(Y_{ik}|s_i, \bm x_i)$ does not depend on $s_i$,
$H_0:\gamma_k =\sigma_{ak}^2=0$ versus $H_A:\gamma_k \neq 0~\text{or}~\sigma_{ak}^2>0.$
\bc Under the null, it is known 
that the usual asymptotic chi-square distribution of the
likelihood ratio (LR) test does not hold. 
The LR test statistic does not follow
a standard chi-square distribution because the value of the variance 
component, $\sigma_{ak}^2$, under the null is on the boundary of the parameter space. 
{For example, for testing a single variance component, 
\cite{self1987asymptotic}
found that under the null,  the LR test
follows a $50:50$ mixture of chi-square distributions (also, see  
\citealp{shapiro1985asymptotic,stram1994variance,stoel2006likelihood,wu2013likelihood}). 
However, when testing multiple variance components, the weights of the mixture distributions cannot
be easily expressed \citep{shapiro1985asymptotic,wu2013likelihood}.
Moreover, for the specific case of testing that the variance component of a penalized spline equals 0, 
\cite{crainiceanu2004likelihood} showed that the LR test
does not follows a $50:50$ mixture of chi-square distributions; the general results of \cite{self1987asymptotic} 
cannot be applied because the shared spline random effects across subjects lead to lack of independence of observations under the alternative.
\cite{crainiceanu2004likelihood} derived the appropriate mixture of chi-square distributions
for a single outcome penalized spline mixed effect model: as the number of subjects $N \rightarrow
\infty,$ the distribution of the  LR statistic for only testing the variance
component equals 0 $(\sigma_{ak}^2=0)$ has an asymptotic distribution that is
a mixture of chi-squares \citep{crainiceanu2004likelihood}
\begin{equation}
(1 - d_N) \chi_0^2 + d_N \chi_1^2 \ ,
\eqx"pvlr"
\end{equation}
where $0 < d_N < 1$ is a function of $N.$  
However, although $d_N$ can be obtained for the case of a univariate outcome,                                                   
the  asymptotic approximation (mixture of chi-squares) often does not perform
well in finite samples \citep{CRAINICEANU200435,lee2012permutation}.
Because of these issues surrounding the asymptotic distribution of the LR test of the variance
component equaling 0 $(\sigma_{ak}^2=0)$, \cite{crainiceanu2002probability}
proposed simulating the finite-sample distribution of the LR statistic under the
posed linear mixed effect model.
In contrast, for the linear mixed model representation of a penalized spline for a univariate         
outcome,  
\cite{lee2012permutation} proposed a permutation test to
calculate a $p$-value for the LR test statistic for the null hypothesis 
that the variance component equals 0 $(\sigma_{ak}^2=0)$; the 
permutation test is formed by permuting 'Cholesky'
transformed residuals within and between subjects.

For penalized splines for multivariate outcomes, and possibly additional
covariates, where we need to jointly test that certain $K$ fixed effects and
$K$ variance components equal 0,  the appropriate mixture of chi-square
distributions for the LR test is not straightforward to obtain and, based on the results of
\cite{CRAINICEANU200435}, can be expected to perform poorly in finite samples. } This has provided the impetus to consider permutation tests as a practical alternative. Specifically, in the next section, we extend the \cite{lee2012permutation} permutation approach to jointly test fixed effects and spline variance components in a multivariate setting with $K$ outcomes,  and
propose two additional permutation tests. Alternatively, a popular approach for
testing for the significance of penalized spline terms is the parametric GAM 
approach proposed by \cite{wood2013p}; this approach is compared to the proposed permutation tests, via simulations, in Section 4. 
\ec

\section{Permutation Tests}

{ Recall that our interest is in the joint test of $H_0:\gamma_k =\sigma_{ak}^2=0,$
for $k=1,...,K$ in the linear mixed model in (\eqn"LMM2"),  i.e.,  that $2K$
parameters equal to 0.  Here we describe our three permutation tests to
calculate a $p$-value of the LR test statistic for this null hypothesis.} { Throughout this section, the three permutation approaches assume that (\ref{LMM2}) is the true model.}   

\subsection{Permuting Covariate}

Our first permutation test is the direct permutation of the quantitative
covariate $s_i.$  { Permuting the covariate was first proposed in \cite{draper1966testing} to test regression coefficients and later in \cite{raz} for an F-test for the linear effect of a quantitative
covariate in a univariate linear regression model.} { Permuting covariates for univariate outcomes was shown
theoretically in \cite{raz} and \cite{diciccio2017robust} to have the correct Type I error rate under the null; see also the review paper by \cite{WINKLER2014381}, which verifies this result using simulation. We follow the approach of permuting the covariate $s_i$ and assume that (\ref{LMM2}) is the true model.
\bc We explore with simulations whether the performance in the multivariate case is similar to that in the univariate case as reported in \cite{WINKLER2014381} when permuting the covariate for the tests of penalized spline models. \ec}

Under the null of no association between $\bfY_i$ and $s_i,$ $s_i$ is simply a
random number assigned to subject $i$.
Specifically, in our permutation test, the $s_i$'s are randomly
permuted, and a random value of $s_i$ is reassigned to each subject. Since the
number of permutations of $s_i$ can be excessively large, we recommend using
Monte Carlo methods to obtain an estimate of the exact permutation \textit{p}-value.

Our proposed general algorithm to obtain a Monte Carlo estimate of the exact
permutation \textit{p}-value of the permutation covariate test is:

\begin{enumerate}
\item \bc Calculate the LR test statistic for $H_0:\gamma_k =\sigma_{ak}^2=0, k=1,...,K$ versus $H_A: \gamma_k\neq 0$ or $\sigma_{ak}^2>0$ for some $k$, in the original sample, and denote this by $LR_{obs}.$\ec

\item Randomly permute the $s_i$, reassigning $s_i$ to subjects, and obtain the maximum likelihood (ML) estimation of the model parameters under the alternative, and then calculate the test statistic $LR$ for the permutation.  

\item  Repeat step 2 a large number, say $M,$  times, which yields $M$ test
statistics,  say $LR^m,$ $m=1,...,M.$

\item   Calculate the \textit{p}-value for the permutation test as the proportion of
permutation samples with $LR^m \ge LR_{obs}.$

\end{enumerate}

{ Note, when calculating the LR statistic under permuted  $s_i$,  we only
need to re-estimate the model parameters and log-likelihood under the
alternative. The model parameters and resulting log-likelihood do not need to
be re-estimated under the null.   This is because under the
null, the permuted covariate does not appear in the model, so the estimated
model parameters and the log-likelihood are the same under the null for any
permutation (as well as under the null in the original dataset before
permuting). Thus, to calculate a \textit{p}-value, one only  needs to calculate
the proportion of permutation samples with the log-likelihood under the
alternative greater than or equal to the log-likelihood under the alternative in
the original sample. As described below, permuting the residuals requires
re-estimation of the parameters (and thus likelihoods) under the null and the
alternative.  Also, note that the permutation test based on permuting the covariate
is not affected if the outcome vector is unbalanced in the sense that not all subjects have all
$K$ outcomes, \bc i.e., when their is imbalance due to incompleteness or missing data in the common set of $K$ outcome variables.  \ec This is because each subject has a single $s_i$, so permuting an
$s_i$ to an different subject having fewer than $K$ outcomes does not create any
problems.  }

\subsection{Permuting Residual Vector}

{  Permuting residuals for testing for the linear effect of a continuous covariate in a univariate linear regression model has been proposed in \cite{freedmanandlane}, and has been shown to have correct Type I error in simulations (e.g., \citealp{andersonandlegendre,WINKLER2014381}).
The review paper by \cite{WINKLER2014381} also illustrated that permuting residuals has comparable Type I error and power to permuting the covariate for the linear effect of a quantitative covariate in a univariate linear regression model. For multivariate/repeated measures data, the residual vectors for subjects must be permuted intact in order to preserve the covariance among the outcomes and ensure asymptotic exchangeability (i.e., the joint distribution of the residual vectors is invariant to the permutation) under the null  when (\ref{LMM2}) is correctly specified. Here, we extend
these approaches to a permutation test for variance components in penalized
splines for multivariate outcomes.

Under the null of no association between $\bfY_i$ and $s_i,$  the linear
mixed models in  (\eqn"LMM1") and (\eqn"LMM2") reduce to 
$$
Y_{ik} = \beta_{0k} + \bfbeta_{1k}'\bfx_i  + e_{ik}, 
$$
which we write in vector notation as
$$
\bfY_i = X_i \bfbeta + \bfe_i
$$
where $\bfe_i=[e_{i1},...,e_{iK}]'$ and $\bfbeta$ has $k^{th}$ row equal to
$[\beta_{0k},\bfbeta_{1k}']$ with corresponding covariate matrix $X_i.$  Here,
$Var(\bfY_i|X_i) = Var(\bfe_i|X_i) =\bfSigma.$   Under the null,  we can
estimate $\bfbeta$ via ML in any linear mixed model program.  We denote the
estimated residual vector as 
\begin{equation}
\hbfe_i = \bfY_i - X_i \hbfbeta \ .
\eqx"residV"
\end{equation}
Under the null,  $\hbfe_i  \sim N(0,\bfSigma).$  Because the estimated residuals
are asymptotically exchangeable,  we can create ``permuted outcomes'' by permuting these
estimated residual vectors and then adding the permuted residual vector for
subject $i$ to the estimated predicted mean $ X_i \hbfbeta$ under the null.}

{ Our proposed general algorithm to obtain a Monte Carlo estimate of the
exact permutation \textit{p}-value of the residual vector permutation  test differs from the algorithm in Section 3.1 only in the second step:

\begin{enumerate}
\item[2.] Randomly permute the $\hbfe_i$, and reassign $\hbfe_i$ to subjects. Denote
subject $i$'s reassigned residual vector as  $\hbfe_i^{*}$.  Create the
permutation outcome vector for subject $i$ as
$$
\bfY_i^{*} = X_i \hbfbeta + \hbfe_i^{*}
$$
Use ML to re-estimate the model parameters and log-likelihoods under
both the
null and alternative, and then calculate the test statistic $LR$ for the
permutation. 
\end{enumerate}
}  
{ If the data are unbalanced, that is, not all subjects have all $K$
outcomes, directly permuting the residual vectors is problematic because a
subject with $K$ outcomes could be reassigned (in a permutation) a residual
vector from a subject with less than $K$ outcomes.  To fix this issue,  one can
instead permute Cholesky residuals within and between subjects (see Section \ref{sec:perm-chol}).}

\subsection{Permuting Cholesky Residuals}
\label{sec:perm-chol}

{ We extend the Cholesky residual permutation approach of \cite{lee2012permutation} to multivariate outcomes \bc to enable permutation of all $N\times K$ residuals, instead of the $K$ residual vectors\ec. As discussed in the previous section,
under the null, when (\ref{LMM2}) is correctly specified, $\hbfe_i  \sim N(0,\bfSigma).$  Let $\hbfSigma$ denote the ML estimate of $\bfSigma$ under the null, and let $S$ denote the Cholesky
decomposition of $\hbfSigma$ i.e. $\hbfSigma=S'S.$ Then, the Cholesky residual
vector, denoted  $\hbfr_i = (S')^{-1}\hbfe_i,$  is  distributed approximately
$\hbfr_i   \sim N(0,I),$  where $I$ is the identity matrix.  Thus, the elements
of   $\hbfr_i$ are approximately independent and exchangeable both within and between subjects.  If we let
the vector $\hbfr' = [\hbfr_1',...,\hbfr_N']$ denote the combined vector of Cholesky
residual vectors across all subjects, we can randomly permute all of the
elements of $\hbfr$ regardless of subject.  With completely balanced data,
there are $N\times K$ elements of $\hbfr$ and thus $(NK)!$ permutations using these 
Cholesky residuals. }

{ Our proposed general algorithm to obtain a Monte Carlo estimate of the
exact permutation \textit{p}-value of the Cholesky residuals permutation  test differs from the algorithm in Section 3.1 only in the second step:

\begin{enumerate}
\item[2.] Randomly permute all elements of the $NK \times 1$ vector $\hbfr$ to give
the permutation vector $\hbfr^{*}.$  For the $i$th subject,  assign the Cholesky
permuted residual vector  $\hbfr_i^{*}$ as the $(K(i-1) + 1),...,Ki$ elements of
$\hbfr^{*}.$  Next,  create $\hbfe_i^{*}= S'\hbfr_i^{*}$. Note,  under the null,
$\hbfe_i^{*}$ will have mean vector 0 and covariance matrix $S'Var(\hbfr_i^{*})S
= S'S = \hbfSigma.$ As with the permutation test of the residual vectors, we now 
create the
permutation outcome vector for subject $i$ as
$$
\bfY_i^{*} = X_i \hbfbeta + \hbfe_i^{*}
$$
Use ML to re-estimate the model parameters and log-likelihoods under
both the
null and alternative, and then calculate the test statistic $LR$ for the
permutation.  
\end{enumerate}

 \bc
Although the elements of the Cholesky transformation of a subject's residual vector are
uncorrelated, in theory they are only independent if the errors are normally
distributed. Thus, it is possible that the Cholesky approach is more sensitive
to non-normality of the errors. We explore the potential sensitivity of the Cholesky approach to non-normality of the errors in simulations studies reported in Section 4.  \ec

With the three permutation approaches,  permuting the covariate should be faster
computationally since we do not have to re-estimate the model under the null.
Further, both permuting the covariate and permutation of Cholesky residuals
do not require balanced data.  }

\section{Simulation Study}

In this section, we use simulation to study the Type I error and power of the three different permutation tests, as well as the parametric test  \citep{wood2013p} as implemented in 
the \texttt{mgcv} R package \citep{wood2015package}. 

\subsection{Details}
We assume that $(X_i,S_i,Y_{ik})$ is generated from the following model:
\begin{equation}
\begin{aligned}
Y_{ik} &= \beta_0 + \beta_1 X_i + \gamma_{k}\sin(2S_i) + e_{ik},\\
(X_i,S_i)&\sim \mathcal N\left(\left(
\begin{array}{c}
     0 \\
     0 
\end{array}\right), \left(
\begin{array}{cc}
     1&0.5  \\
     0.5&1 
\end{array}\right)\right).
 \end{aligned}
\eqx"sm"
\end{equation}


\noindent In this simulation study, we further assume that $e_{ik} = b_i + \epsilon_{ik}$ where $b_i\sim \mathcal N(0,\rho)$ is a random intercept and $\epsilon_{ik}$ with $Var(\epsilon_{ik}) = 1-\rho$ is the within-subject random error. 
\bc Specifying $\mathbb E(Y_{ik}|X_i,S_i)$ to depend on $\sin(2S_i)$ is a common choice in the literature for penalized spline regression \citep{wood2003thin,chen2011penalized,chen2013marginal}. Accurate estimation of sinusoidal functions requires an approach beyond regular polynomial regressions and thus the need for a flexible approach such as penalized spline models.
\ec 
We set $\beta_0 = 0$ and $\beta_1 = 1$. This model specification implies that the multiple outcomes have marginal variance $Var(Y_{ik}|X_i,S_i)= Var(b_i) + Var(\epsilon_{ik})=1$ and compound symmetric correlation  $\rho=Corr(Y_{ij},Y_{ik}|X_i,S_i)$ for $j \neq k.$ { Although the correlation
among the multiple outcomes is likely to be more complicated (e.g., unstructured as recommended in Sec. 2) than compound symmetric
in many applications, our motivation for the use of a compound symmetric correlation in the simulations is to make it more transparent how
the tests perform under a single parameter describing lower and higher correlations.  We fix the marginal variance of $Y_{ik}$ given $(X_i,S_i)$ to be 1 so that varying $\rho$ can be done by varying $Var(b_i)=\rho$. This formulation of the linear mixed model in (\eqn"sm") \bc with $e_{ik} = b_i + \epsilon_{ik}$ \ec makes it easier to specify a non-normal distribution for $Y_{ik}$ given $(X_i,S_i)$ with compound symmetric correlation, as we describe below.  }

In the simulation, we study the properties of the tests with different configurations of $N$, $K$, $\rho$ and { regression coefficients of $\sin(S_i)$ for the $K$ outcomes, } ${\bf \gamma} = (\gamma_1,\ldots,\gamma_K)$. Specifically, we let $N\in\{50,100,200\}$, $K\in\{5,10\}$, $\rho = \{0.25,0.75\}$; \bc the choice of values for $N$ and $K$ was motivated by the dimension of the application dataset from the CCTS trial. \ec
We consider three different values of $\bm\gamma = (\gamma_1,\ldots,\gamma_K)$:
\begin{enumerate}
    \item (\textbf{Sparse}) $\gamma_1 = 0.3,\gamma_{k}=0, k=2,\ldots,K$,
    \item (\textbf{Non-uniform}) $\gamma_k = 0.5/k$, $k = 1,\ldots, K$,
    \item (\textbf{Uniform}) $\gamma_k = 0.75$, $k=1,\ldots,K$.
    \end{enumerate}
 These three scenarios aim to capture distinct representative relationships between $Y$ and $S$. \bc The sparse and uniform specifications aim to examine the performance of a test in two extreme cases, where the association is only present in one of the outcomes (sparse) and the association is uniformly strong (uniform). Non-uniform $\gamma$ serves as a middle ground and, arguably, may also be more likely to be the case in real applications. \ec Through trial and error in small scale simulations,  we choose the values of the non-zero $\gamma$ coefficients so that the estimated power for any of the tests is bounded away from 0 and 1.   \bc We let $b_i\sim \mathcal N(0,\rho)$ for all simulations \ec and consider two different distributions of the within-subject random error, where $\epsilon_{ik}\sim \mathcal N(0,1-\rho)$ and $\epsilon_{ik}\sim \sqrt{1-\rho}\cdot\text{SLN}(0,1)$. Here $\text{SLN}(0,1)$ is the lognormal distribution $\text{LN}(0,1)$ { after subtracting off its mean $\exp(1/2)$ and scaling by its standard deviation $\sqrt{(\exp(1)-1)\exp(1)}$,} so that $\text{SLN}(0,1)$ has mean 0 and variance 1.  Note, for both distributions of $\epsilon_{ik},$ the compound symmetric correlation $\rho=\text{Corr}(Y_{ij},Y_{ik}|X_i,S_i)$ still holds, but we use the skewed SLN to examine the robustness of the tests against model misspecification (non-normality of $Y_{ik}$).  Further,  exchangeability of the residual vectors still hold for this non-normal distribution, so that we would expect the permutation tests to perform well with respect to Type I error.


For each configuration of the model parameters we performed 2000 simulation replications to estimate the Type I error and 1000 simulation replications for
the power. \bc 
The increased number of replications when examining the Type I error was to ensure adequate precision when estimating a probability that is close to 0, thereby obtaining error bands that are relatively narrow with a Monte Carlo standard error of the estimated probability less than 0.005.  
\ec As suggested in \cite{manly2018randomization}, we used $M=1000$ permutation samples in
our simulations. The estimated Type I error under the
null,  and power under the alternatives, were calculated as the proportion of
the simulation replications in which a given \textit{p}-value was less than
0.05. We leave out the configurations where $N=200$ and $K=10$ due to computational cost;  { even without  configurations with $N=200$ and $K=10$ , we are able to compare \bc three different sample sizes  $N\in\{50,100,200\}$ when $K=5$ and two different number of outcomes $K\in\{5,10\}$ when $N\in\{50,100\}.$ \ec }

{ For all permutation tests, we use equally spaced  knots between -2 and 2 to generate the piece-wise linear spline basis, as relatively few values of $S$ are expected outside of this range.} The properties of penalized splines suggest that inference should be relatively insensitive to the number of knots chosen \citep{ruppert2003semiparametric}.   We performed a small scale simulation (see online appendix) and showed that the Type I error and power were not very sensitive when we chose $C=10, 20, 30, 40, 50$ knots. Thus we only report results with $C=30$. For the parametric GAM test, we penalized the first derivative of the smoothing term of $S$ and select the number of knots by optimizing the GCV metric. \bc This selection step is an integral part of the GAM model fitting procedure \citep{wood2015package}\ec.  Across all configurations, the optimal number of knots for the parametric GAM test varies between 5 and 10. The parametric GAM test produces  one \textit{p}-value for each outcome separately when testing the association between $Y_{ik}$ and $S_i$ across $k=1,\ldots,K$. We used the Bonferroni corrected \textit{p}-value derived from these $K$ \textit{p}-values as the result of the joint test. We note that we did not use the joint test of multivariate outcomes provided in the \texttt{mgcv} R package based on a generalized likelihood ratio (GLR) test since it has been shown to have inflated Type I error \citep{SCHEIPL20083283} when the comparisons involve penalized terms; preliminary simulations also confirmed that the Type I error was inflated in our setting.  Specifically, for our simulation scenarios where $\epsilon_{ik}$ follows a normal distribution, the range of the Type I error of the GLR test was found to be between 0.2 and 0.3, while when $\epsilon_{ik}$ follows a SLN distribution, the range was found to be between 0.4 and 0.6.



\subsection{Results}
The simulation results are presented in Table \ref{tab:normal} for normally distributed errors and in Table \ref{tab:sln} for SLN distributed errors. \bc Because the pattern of results for $N=50$ is similar to that for $N=100$, we present results for $N=50$ in supplementary tables\ec. We see that the estimated  Type I error from the simulations for all 3 permutation tests are close to 0.05 with 95\% Wald confidence intervals for the Type I errors covering 0.05 (see online appendix). The Type I error for the parametric GAM test is slightly elevated in this case. However, when $\epsilon_{ik}$ follows SLN, the  Type I error for permuting Cholesky covariates is slightly larger than 0.05 whereas the parametric GAM test fails to control Type I error. On the other hand, both permuting covariates and residual vectors have well controlled Type I error. The results confirm the robustness of permutation tests when applied to non-normal data. Permuting Cholesky residuals tends to be less robust than the other two permutation approaches, possibly due to its reliance on the normality assumption to guarantee the independence of transformed residuals. The parametric GAM test seems to not be applicable when data are not normally distributed.

When $\epsilon_{ik}$ is normally distributed, the parametric GAM test generally has the highest power among all four tests, except for the case of uniform $\bm\gamma$ (only for high $\rho$ when $N=100$ and $N=200$ and for both high and low $\rho$ when $N = 50$). The three permutation tests have nearly identical power across the different configurations. As expected, increasing $N$ boosts the power consistently across all configurations while increasing $K$ typically has minimal or even adverse effect on power (e.g., uniform $\bm\gamma$). Increasingly $\rho$ has a positive impact on power for all four tests when $\bm\gamma$ is sparse and non-uniform. When $\bm\gamma$ is uniform, larger $\rho$ leads to smaller power. This decrease in power for a uniform effect with increasing correlation agrees with similar simulation results found in the literature \citep{yoon2011alternative, bubeliny2010hotelling}. 

When $\epsilon_{ik}$ follows SLN, the power of the parametric GAM test cannot be compared to the others due to its lack of control of Type I error. All permutation tests have decreased power compared to the case where the error is normally distributed. Permuting covariates and residual vectors still have nearly identical power across all scenarios whereas permuting Cholesky residuals tends to have a slightly higher power at the expense of elevated Type I error. The effects of varying $N$, $K$, $\rho$, and $\bm\gamma$ remain the same as in the normally distributed random error case.

Since the parametric GAM test cannot control Type I error when applied to non-normal data, permutation tests may be preferred in data application when the normality assumption is not satisfied or is difficult to verify. Since all three permutation tests have similar power and permutation of Cholesky residual tends to be less robust against non-normal errors, we should prioritize permuting of residual vectors or permuting of covariates over it. We also observe that permuting covariates generally is more computational efficient than permuting residual vectors since there is no need to refit the null model for each permutation replicate. Therefore, permuting covariates may be preferred when computational resources are limited.

\section{Application to Collaborative Cocaine Treatment Study}


Our motivating application is from a longitudinal clinical trial, the National
Institute on Drug Abuse Collaborative Cocaine Treatment Study (CCTS)
\citep{crits1999psychosocial}, where there is scientific interest in 6 month
post-treatment changes in 5 psychosocial problem domains. Thus, the outcomes are
the simple changes from baseline in legal, employment, family, psychological and
medical problems at 6 months of follow-up beyond the end of treatment. These 5
non-substance related problem domains, as assessed by the Addiction Severity
Index (ASI; \citealp{mclellan1992fifth}), have significant societal consequence. Using data from the CCTS, we
are interested in the association between within-treatment frequency of drug use
(based on urine toxicology screens) and the changes from baseline to 6 month
post-treatment (12-month post-baseline) follow-up in these 5 problem domains. 

Briefly, the NIDA CCTS was a multisite clinical trial of patients randomized to
4 psychosocial treatments for 6 months: group drug counseling (GDC) alone,
individual cognitive therapy (CT) plus GDC, individual supportive-expressive
(SE) psychodynamic therapy plus GDC, and individual drug counseling (IDC) plus
GDC (see \cite{crits1999psychosocial} for additional details). In this trial, N=487 participants were randomized to one of the 4 treatment
groups; participants were 18 years of age or older (mean age 33.9 yrs), 23\%
female, 58\% white, and met criteria for current cocaine dependence according to
Diagnostic and Statistical Manual of Mental Disorders, Fourth Edition \citep{association2000diagnostic}. 
In terms of within-treatment substance use, 
a composite cocaine use measure, constructed by pooling information from
self-report data and weekly observed urine samples, was used to code each month
of treatment as abstinent versus any cocaine use. Thus, in the CCTS there were 6
binary monthly assessments of cocaine use, one for each of the 6 months of
active treatment. Within-treatment frequency of use was defined as the
proportion of positive monthly urine toxicology screens.

To assess the association between within-treatment frequency of use and
improvements in post-treatment follow-up assessments of the ASI problem domains
at 6  months post-treatment, we used a multivariate penalized piecewise-linear
regression model.  With changes in the 5 problem domain scores as the outcomes,
$Y_{ik}$ for $k=1,...,5$, we modeled within-treatment frequency of use, $s_i$,
as a penalized piecewise-linear spline with $5$ knots located at $\kappa_1=1/6,
\kappa_2=1/3, \kappa_3=1/2, \kappa_4=2/3, \kappa_5=5/6:$  
$$
Y_{ik} = \beta_{0k} + \bfbeta_{1}'\bfx_i + \gamma_k s_i + \sum_{c=1}^5 a_{ck}
(s_i-\kappa_c)_{+}
+ e_{ik},
$$
where 
$\bfx_i$ 
included indicator variables for study site and treatment group, and also the
baseline assessment of the ASI problem domains. The random effects for the spline 
$a_{1k},...,a_{5k}$ are assumed to have independent normal distributions,
$a_{ck} \sim N(0,\sigma_{ak}^2)$, $c=1,\ldots,5$. Finally, we assumed that the covariance matrix
is unstructured, $\var(\bfY_i|\bfx_i,s_i) = \bfSigma$ with
$\var(Y_{ik}|\bfx_i,s_i) = Var(e_{ik}) = \sigma_k^2$ and
$\text{cov}(Y_{ij},Y_{ik}|\bfx_i,s_i) = \sigma_{jk}.$ This unconstrained model allows
for completely different functional relationships between within-treatment
frequency of use and each of the five problem domains; under the null $H_0:\gamma_k
=\sigma_{ak}^2=0,$ $k=1,...,5,$  there are 10 parameters set to 0. { We
used $M=10,000$ permutation replicates for all of our tests. Because some patients
have missing $Y_{ik}$'s,  we did not perform the residual vector permutation
test, but only the covariate permutation and Cholesky residual permutation
tests. We compared these to the the parametric GAM test where B-splines with five internal knots were used and the first derivative of the smooth term is penalized.}

\begin{figure}
\centering
\includegraphics[width=0.95\linewidth]{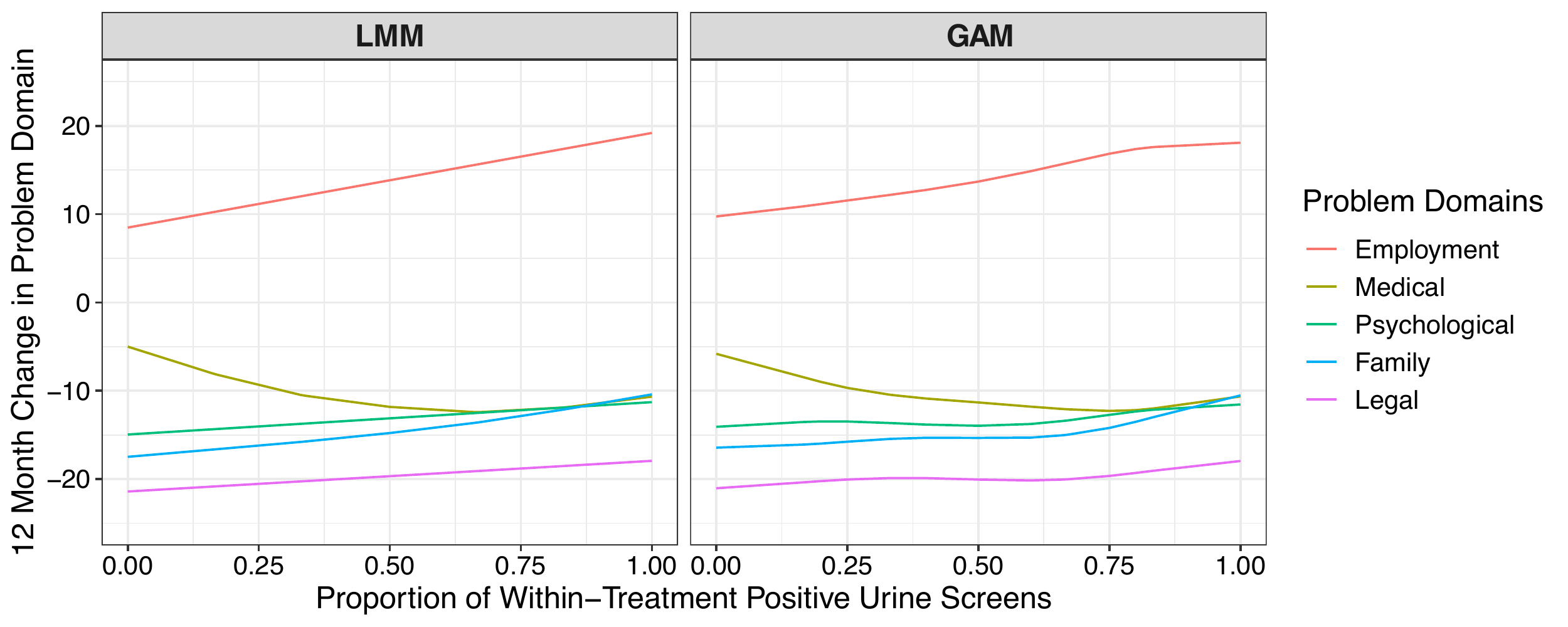}
\caption{Fitted curves for the relationships between  
within-treatment frequency of cocaine use and changes in the 5 problem domains
based on the unconstrained proposed model (LMM) and GAM.}
\label{fig:app-res}
\end{figure}

Results of the { 2 permutations tests as well as the parametric GAM test } for the changes in the five ASI
problem domains at 12 months post-baseline are presented in Table \ref{tab:app-res}. { For the joint tests, the
covariate permutation \textit{p}-value $=0.0008$ and the Cholesky residual permutation \textit{p}-value $=0.0015$; the parametric GAM \textit{p}-value $=0.0042$.}  Thus, the joint tests suggest there is evidence that
within-treatment frequency of cocaine use is associated with the problem domains
at 12 months post-baseline. The joint permutation tests are omnibus tests and do not
indicate which of the 5 problem domains are associated with within-treatment
frequency of cocaine use. { We therefore performed univariate tests by permuting the covariates or residuals to evaluate domain-specific associations with within-treatment frequency of cocaine use. The results are also collected in Table \ref{tab:app-res}. The two permutation tests give similar univariate \textit{p}-values for all outcomes and both approaches suggest that employment and family problems are the two domains that are significantly associated with the frequency of cocaine use (Bonferroni corrected residual permutation \textit{p}-values of 0.03 and 0.002 respectively). Similar results are also observed when the parametric GAM test was used. 

We also visualized the fitted curves for all five problem domains as functions of within-treatment frequency of cocaine from our LMM approach (Fig. \ref{fig:app-res}, left) as well as from GAM (Fig. \ref{fig:app-res}, right). 
These curves were generated at the fixed reference levels of treatment group and study site while the values of baseline assessment of problem domains were all fixed at the mean across all domains and all subjects.
Note that the resulting curves from LMM approach were based on REML, instead of ML, fit to the mixed effects model; \bc although ML must be used for constructing LR test, REML can be used for estimation of the fitted curves. \ec  The estimated smoothing curves of each problem domain from both approaches are very similar. The curves from GAM tend to be more nonlinear. For the two domains (Employment and Family) with statistically significant associations with within-treatment frequency of cocaine use, both plots  indicate that a 50\%
difference in within-treatment frequency of cocaine use, say 75\% use versus
25\% use, is associated with an approximate 5.5 point mean increase in employment problems and 2.5 point mean increase in family problems.
}


\section{Discussion}
\bc
In many applications, the relationship between the mean of an outcome and a quantitative covariate is complex and cannot be described by simple parametric functions (e.g., polynomial trends). In these settings, flexible nonlinear relationships can be incorporated using penalized splines. In this paper we have focused on the application of penalized splines in joint models that allow flexible nonlinear relationships for multiple outcomes. 
Due to the additional complexity of multiple correlated outcomes, the fitting of joint penalized spline models is potentially very challenging. We overcome this challenge by exploiting the known connection between penalized spline models and linear mixed effects models (e.g., \citealp{ruppert2003semiparametric}). The linear mixed effects model representation simplifies model fitting in the multivariate outcome setting, and can also be implemented in existing statistical software, while allowing flexible nonlinear relationships with each outcome and properly accounting for the correlation among the outcomes. However, statistical inference based on 
a joint test of zero-effect of the quantitative covariate on the multiple outcomes involves testing that the 
variance components associated with each outcome are jointly equal to 0. When testing that a variance component equals 0, it is well-known that the usual asymptotic chi-square distribution of the
likelihood ratio (LR) test under the null does not hold \citep{miller1977asymptotic,lin1997variance,verbeke2003use}. To overcome this limitation, we have proposed three permutation tests for the likelihood ratio test statistic; one based on permuting the quantitative covariate, the other two based on permuting residuals.
Thus, the main contribution of this paper to methodology is that we have proposed an extension of the permutation approach to jointly test flexible nonlinear relationships, incorporated using penalized splines, 
to the {\it multivariate} setting with multiple outcomes. This includes a natural extension of the Cholesky residual permutation approach of \cite{lee2012permutation} to multivariate outcomes.
\ec

Results from the simulation study \bc lead to the following general recommendation concerning the three permutation tests we have proposed. 
A permutation test, based on permuting either the covariate or the vector of residuals, provides a robust alternative to the commonly used parametric GAM test proposed by \cite{wood2013p}, albeit with lower power for many alternatives where the parametric test appears valid. We note that for the case of non-normal errors, the permutation test based on Cholesky residuals had inflated Type I error rate and seemed less robust to non-normal errors. Similarly, the parametric test was found to have highly inflated Type I error rate with non-normal errors, suggesting this test is very sensitive to violations of distributional assumptions.\ec

Finally, we note that the presentation of penalized splines, and our proposed
permutation tests, has been for the special case of a simple piecewise linear
spline. This focus on piecewise linear splines was primarily for notational
convenience; all of the proposed tests generalize in a natural way to more
flexible piecewise polynomial response models using alternative basis functions
such as cubic splines and B-splines. For example, for cubic splines, the
proposed $2K$ { parameter} joint tests simply become $4K$ parameter tests.
The proposed permutation tests are not any more difficult to implement when
other basis functions have been adopted. In addition, the focus of this paper
has been on linear models for $K$ quantitative or continuous outcomes. However,
we note that the proposed permutation tests can equally be applied in penalized
splines for generalized linear models, e.g., logistic regression models. That
is, the permutation tests can be implemented using a generalized linear mixed
model (GLMM) representation of penalized splines; this is a topic of further
research. In closing, we note that specification of penalized splines for
multiple outcomes using the linear mixed effects model representation is
straightforward to implement in widely available software for fitting linear
mixed effects models. Obtaining permutation \textit{p}-values for the proposed joint
tests requires only modest additional programming; an R function   for implementing
the proposed permutation tests is included in the online appendix.

\bigskip
\begin{center}
{\large\bf SUPPLEMENTARY MATERIAL}
\end{center}

\noindent A written supplementary material contains additional results on Type I error and power in simulations, illustrating confidence intervals and impact of number of knots $C$, {\bc and for smaller sample size $N = 50$\ec}. The R code for simulations and data application is provided as a separate zip file.

\bibliographystyle{apacite}
\bibliography{reference}

\newpage

\begin{table}[htbp]
  \centering
\caption{Results for Type I error (Null) and power from simulations when $\epsilon_{ik}\sim\sqrt{1-\rho}\cdot\mathcal N(0,1)$ in equation (\eqn"sm"). GAM stands for the parametric GAM test.}

    \resizebox{\textwidth}{!}{\begin{tabular}{c|c|c|c|c|c|c|c|c|c|}
    \multirow{2}[3]{*}{$(N,K)$} & \multirow{2}[3]{*}{Methods} & \multicolumn{4}{c|}{$\rho=0.25$} & \multicolumn{4}{c|}{$\rho = 0.75$} \bigstrut[b]\\
\cline{3-10}          &       & Null  & Sparse & Non-unif & Uniform & Null  & Sparse & Non-unif & Uniform \bigstrut\\
    \hline
    \multirow{4}[2]{*}{$(200,5)$} & Residual & .053 & .152 & .469 & .707 & .051 & .712 & .926 & .437 \bigstrut[t]\\
          & Covariate & .056 & .153 & .460 & .701 & .053 & .713 & .927 & .431 \\
          & Cholesky & .053 & .149 & .466 & .713 & .050 & .717 & .927 & .437 \\
          & GAM  & .064 & .554 & .918 & .965 & .073 & .976 & .997 & .181 \bigstrut[b]\\
    \hline
    \multirow{4}[1]{*}{$(100,5)$} & Residual & .049 & .096 & .155 & .386 & .050 & .271 & .428 & .244 \bigstrut[t]\\
          & Covariate & .051 & .091 & .157 & .383 & .050 & .271 & .430 & .247 \\
          & Cholesky & .049 & .096 & .151 & .381 & .052 & .274 & .438 & .242 \\
          & GAM  & .057 & .258 & .547 & .616 & .075 & .700 & .887 & .140 \bigstrut[b]\\
 \hline
    \multirow{4}[2]{*}{$(100,10)$} & Residual & .039 & .089 & .188 & .280 & .052 & .229 & .635 & .182 \bigstrut[t]\\
          & Covariate & .039 & .089 & .192 & .278 & .055 & .232 & .633 & .186 \\
          & Cholesky & .039 & .092 & .188 & .280 & .055 & .229 & .633 & .173 \\
          & GAM  & .064 & .206 & .520 & .210 & .070 & .704 & .968 & .093 
    \end{tabular}}%
  \label{tab:normal}%
\end{table}%

\begin{table}[htbp]
  \centering
  \caption{Results for Type I error (Null) and power from simulations when $\epsilon_{ik}\sim \sqrt{1-\rho}\cdot\text{SLN}(0,1)$ in equation (\eqn"sm"). GAM stands for the parametric GAM test.}
    \resizebox{\textwidth}{!}{\begin{tabular}{c|c|c|c|c|c|c|c|c|c|}
    \multirow{2}[3]{*}{$(N,K)$} & \multirow{2}[3]{*}{Methods} & \multicolumn{4}{c|}{$\rho=0.25$} & \multicolumn{4}{c|}{$\rho = 0.75$} \bigstrut[b]\\
\cline{3-10}          &       & Null  & Sparse & Non-unif & Uniform & Null  & Sparse & Non-unif & \multicolumn{1}{c|}{Uniform} \bigstrut\\
    \hline
    \multirow{4}[2]{*}{$(200,5)$} & Residual & .052 & .141 & .372 & .597 & .052 & .633 & .816 & .319 \bigstrut[t]\\
          & Covariate & .054 & .139 & .374 & .586 & .054 & .629 & .812 & .311 \\
          & Cholesky & .061 & .149 & .390 & .648 & .067 & .687 & .865 & .388 \\
          & GAM  & .193 & .622 & .918 & .921 & .186 & .979 & .996 & .264 \bigstrut[b]\\
 \hline
    \multirow{4}[1]{*}{$(100,5)$} & Residual & .053 & .075 & .119 & .229 & .052 & .262 & .411 & .151 \bigstrut[t]\\
          & Covariate & .051 & .074 & .120 & .234 & .053 & .259 & .409 & .145 \\
          & Cholesky & .064 & .085 & .137 & .287 & .073 & .307 & .471 & .196 \\
          & GAM  & .215 & .391 & .641 & .561 & .199 & .774 & .909 & .213 \bigstrut[b]\\
    \hline
    \multirow{4}[2]{*}{$(100,10)$} & Residual & .051 & .074 & .089 & .097 & .047 & .153 & .368 & .068 \bigstrut[t]\\
          & Covariate & .051 & .066 & .089 & .083 & .050 & .161 & .368 & .070 \\
          & Cholesky & .063 & .087 & .124 & .147 & .092 & .228 & .493 & .150 \\
          & GAM  & .337 & .474 & .692 & .453 & .341 & .812 & .962 & .382 \\
   
    \end{tabular}}%
  \label{tab:sln}%
\end{table}%


\begin{table}[htbp]
  \centering
  \caption{Univariate and joint tests of associations between problem domains and frequency of cocaine use. The \textit{p}-values of univariate tests are not corrected for multiple comparisons. We only report results from the Cholesky residual permutation approach for the residual-based joint permutation test. For univariate tests, permuting Cholesky residuals is equivalent to permuting residuals directly. For the parametric GAM test, we use the Bonferroni corrected smallest univariate \textit{p}-value as the \textit{p}-value of the joint test.}
    \begin{tabular}{l|c|c|c}
    \multirow{2}[3]{*}{} & \multicolumn{2}{c|}{Permutation} & \multirow{2}[3]{*}{GAM} \bigstrut[b]\\
\cline{2-3}          & Covariate & Residual &  \bigstrut\\
    \hline
    Employment & .0077 & .0051 & .0008 \bigstrut[t]\\
    Medical & .2017 & .1972 & .0743 \\
    Psychological & .1135 & .1164 & .2995 \\
    Family & .0004 & .0004 & .0078 \\
    Legal & .0639 & .0674 & .1472 \bigstrut[b]\\
    \hline
    Joint & .0008 & .0015 & .0042 \bigstrut[t]\\
    \end{tabular}%
  \label{tab:app-res}%
\end{table}%

\newpage

\end{document}